\documentclass[twocolumn,aps,prd,epsf,showpacs]{revtex4-1}
\pdfoutput=1
\usepackage[pdftex]{graphicx}
\usepackage{color}
\usepackage{amsmath}
\usepackage{amsfonts}
\usepackage{amssymb}
\usepackage{dsfont}
\usepackage[]{units}
\usepackage{braket}
\usepackage[utf8]{inputenc}
\usepackage[T1]{fontenc}
\usepackage{url}
\usepackage[english]{babel}
\usepackage{graphicx}
\usepackage{caption}
\usepackage{subcaption}
\usepackage[export]{adjustbox}
\usepackage{dcolumn}
\usepackage{bm}
\usepackage{subfig}
\usepackage{float}
\usepackage{url}
\usepackage{verbatim}
\usepackage[]{cleveref}

\newcommand{\be}{\begin{equation}}
\newcommand{\ee}{\end{equation}}
\newcommand{\bea}{\begin{eqnarray}}
\newcommand{\eea}{\end{eqnarray}}

\newcommand{\bie}{\begin{itemize}}

\newcommand{\eie}{\end{itemize}}
\newcommand{\ben}{\begin{enumerate}}
\newcommand{\een}{\end{enumerate}}

\DeclareMathOperator{\diag}{diag}
\DeclareMathOperator{\Tr}{Tr}

\graphicspath{{figures/}}

\newcommand{\gtapprox}{\raisebox{-0.5ex}{$\,\stackrel{>}{\scriptstyle\sim}\,$}}
\newcommand{\ltapprox}{\raisebox{-0.5ex}{$\,\stackrel{<}{\scriptstyle\sim}\,$}}

\begin{document}

\title{Including heavy spin effects in the prediction of a $\bar{b} \bar{b} u d$ tetraquark with lattice QCD potentials}

\author{$^{(1)}$Pedro Bicudo}
\email{bicudo@tecnico.ulisboa.pt}

\author{$^{(2)}$Jonas Scheunert}
\email{scheunert@th.physik.uni-frankfurt.de}

\author{$^{(2)}$Marc Wagner}
\email{mwagner@th.physik.uni-frankfurt.de}

\affiliation{\vspace{0.1cm}$^{(1)}$CFTP, Dep.\ F\'{\i}sica, Instituto Superior T\'ecnico, Universidade de Lisboa, Av.\ Rovisco Pais, 1049-001 Lisboa, Portugal}

\affiliation{\vspace{0.1cm}$^{(2)}$Johann Wolfgang Goethe-Universit\"at Frankfurt am Main, Institut f\"ur Theoretische Physik, Max-von-Laue-Stra{\ss}e 1, D-60438 Frankfurt am Main, Germany}

\begin{abstract}
We investigate spin effects in four-quark systems consisting of two heavy anti-bottom quarks and two light up/down quarks. To this end we use the Born-Oppenheimer approximation. We utilize potentials of two static antiquarks in the presence of two quarks of finite mass computed via lattice QCD and solve a coupled-channel Schrödinger equation for the anti-bottom-anti-bottom separation. Without taking heavy quark spins into account this approach predicted a $u d \bar b \bar b$ tetraquark bound state with quantum numbers $I(J^P) = 0(1^+)$. We now extend this Born-Oppenheimer approach with coupled channel Schrödinger equations allowing us to incorporate effects due to the heavy $\bar b$ spins. We confirm the existence of the $u d \bar b \bar b$ tetraquark.
\end{abstract}


\maketitle


\section{\label{sec:intro}Introduction}

A long standing problem of QCD is the search for exotic hadrons, beyond the simple quark-antiquark mesons and three-quark baryons \cite{Jaffe:1976ig}. However, studying exotic hadrons, e.g.\ glueballs, hybrid mesons, tetraquarks, pentaquarks or hexaquarks, theoretically turned out to be much harder than expected (cf.\ e.g.\ \cite{Miyazawa:1979vx,Miyazawa:1980ft,Oka:1984yx,Oka:1985vg,Lenz:1985jk,Karliner:2003dt,Heupel:2012ua,Eichmann:2015cra,Bicudo:2015bra,Rupp:2016jdk}). Also experimentally this is a difficult problem, since exotic candidates are typically resonances immersed in the excited hadron spectra, which quickly decay to several non-exotic hadrons. So far only the recently observed tetraquark candidates $Z_c$ and $Z_b$ have survived the scrutiny of the scientific community.

In lattice QCD the study of exotics is extremely challenging. For instance in \cite{Ikeda:2013vwa,Prelovsek:2014swa,Leskovec:2014gxa} the authors searched for evidence of a large tetraquark component in the resonance $Z_c(3940)^-$. To this end large correlation matrices of two-quark and four-quark hadron creation operators (including e.g.\ two-meson and diquark-antidiquark structures) were implemented and investigated. The main difficulty is that the $Z_c(3940)^-$ is a resonance well above threshold with many lighter two-meson scattering states below. No robust evidence of a $Z_c(3940)^-$ tetraquark resonance was found. 

In quark model calculations the issue is as well not settled. Using the perturbative approximation of the resonating group method a preliminary estimation of the partial decay width of the $Z_c(4430)^-$ resonance was found, similar to the one measured by LHCb \cite{Cardoso:2008dd}. However, because tetraquarks are always coupled to meson-meson systems, more sophisticated quark models like the string flip-flop potential for the meson-meson interaction were developed, to solve the problem of Van der Waals forces produced by the two-body confining potentials \cite{Miyazawa:1979vx,Miyazawa:1980ft,Oka:1984yx,Oka:1985vg,Lenz:1985jk,Karliner:2003dt}. However, there is a recent claim that the string flip-flop potentials still produce excessive binding \cite{Bicudo:2015bra}.

Our main motivation is to investigate in detail exotic tetraquarks and mesonic molecules by combining lattice QCD and techniques from quantum mechanics. In this way we avoid to a large extent both the difficulties found in pure lattice QCD computations and in the model dependence of quark model computations.

\begin{figure}[htb]
\centerline{\includegraphics[width=0.95\columnwidth]{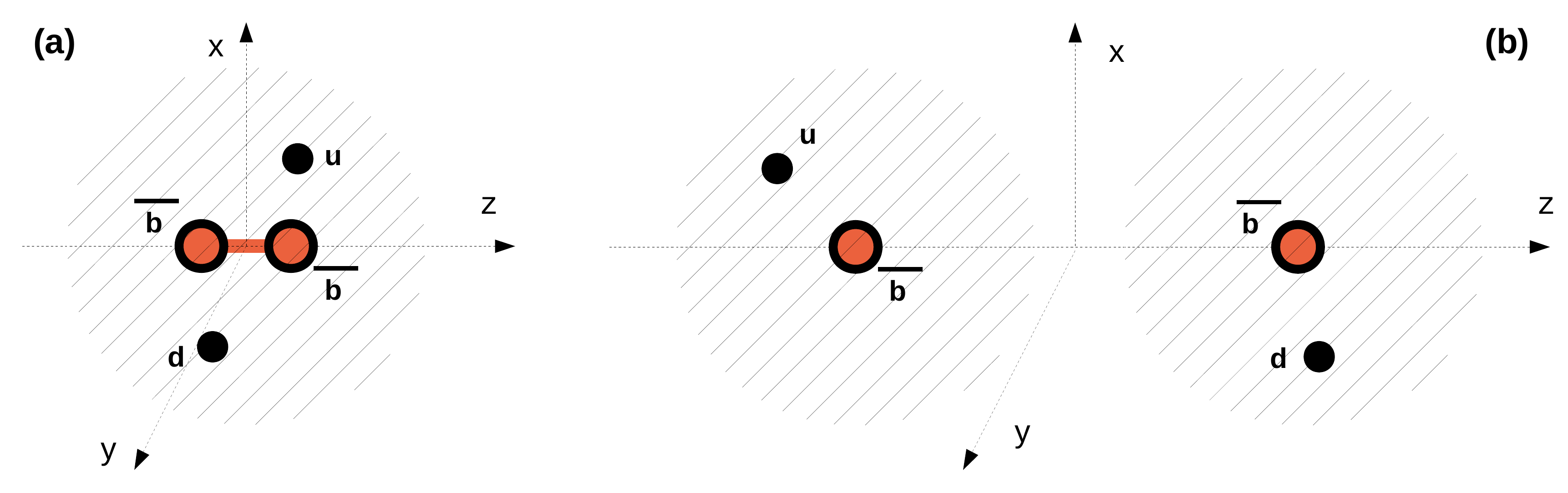}}
\caption{(a)~At very short $\bar b \bar b$ separations the $\bar{b}$ quarks interact with a perturbative one-gluon-exchange Coulomb-like potential. (b)~At large $\bar b \bar b$ separations the light quarks $u d$ screen the interaction and the four quarks form two rather weakly interacting $B$ and/or $B^*$ mesons.
\label{fig:screening}}
\end{figure}

We specialize in systems containing two heavy antiquarks and two lighter quarks. From basic principles of QCD it is clear that such a system, for instance $u d \bar b \bar b$, should form a boundstate, i.e.\ a tetraquark, if the $\bar b$ quarks are very heavy \cite{Ader:1981db,Heller:1986bt,Carlson:1987hh,Lipkin:1986dw,Brink:1998as,Gelman:2002wf,Vijande:2003ki,Janc:2004qn,Cohen:2006jg,Vijande:2007ix}.  To understand the binding mechanism, it is convenient to use the Born-Oppenheimer perspective \cite{Born:1927}, where the wavefunction of the two heavy antiquarks is determined considering an effective potential obtained via a lattice QCD computation of the light quarks. QCD with light quarks and gluons has a characteristic scale of the order of $400 \, \textrm{MeV} \sim 0.5 \, \textrm{fm}^{-1}$, present for instance in the constituent quark mass and in the confinement string tension $\sqrt \sigma$.
At much shorter $\bar b \bar b $ separations $r \ll \sigma^{-1/2}$, the $\bar{b}$ quarks interact with a perturbative one-gluon-exchange Coulomb-like potential. At large separations the light quarks screen the interaction and the four quarks form two rather weakly interacting $B$ and/or $B^\ast$ mesons as illustrated in \Cref {fig:screening}. Thus a screened Coulomb potential is expected. This potential clearly produces a boundstate, providing the antiquarks $\bar b \bar b$ are sufficiently heavy. 

Moreover this class of tetraquarks is related to the doubly-heavy class of closed charm tetraquarks $Z_c^\pm$ and closed bottom $Z_b^\pm$ tetraquarks. The $Z_b^\pm$ was claimed by the BELLE collaboration \cite{Belle:2011aa}, while the $Z_c^\pm$ has received a series of experimental observations by the BELLE collaboration \cite{Liu:2013dau,Chilikin:2014bkk}, the Cleo-C collaboration \cite{Xiao:2013iha}, the
BESIII collaboration \cite{Ablikim:2013mio,Ablikim:2013emm,Ablikim:2013wzq,Ablikim:2013xfr,Ablikim:2014dxl} and the LHCb
collaboration \cite{Aaij:2014jqa}. However, this second class of double-heavy tetraquarks is more difficult to address theoretically, since the  $Z_c^\pm$ and $Z_b^\pm$ are QCD resonances, but not QCD bound states. Thus, we leave it for future studies (cf.\ \cite{Peters:2016isf} for first crude results).

Here we continue our previous studies, where we computed potentials of two static antiquarks in the presence of two quarks of finite mass using lattice QCD. Details on the computation of these potentials can be found in Refs.\ \cite{Wagner:2010ad,Wagner:2011ev,Bicudo:2015kna}. The existence of bound four-quark states, i.e.\ of tetraquarks, is then investigated using the Born-Oppenheimer approximation \cite{Born:1927} and standard techniques from quantum mechanics. Recently, this approach provided evidence for the existence of a $u d \bar b \bar b$ tetraquark with quantum numbers $I(J^P) = 0(1^+)$ \cite{Bicudo:2012qt,Brown:2012tm,Bicudo:2015kna}, while for similar combinations of heavier flavors $s s \bar b \bar b$ and $c c \bar b \bar b$ no bound states seem to exist \cite{Bicudo:2015vta,Peters:2015tra} (the latter is consistent with lattice QCD computations considering four quarks of finite mass; cf.\ e.g.\ \cite{Ikeda:2013vwa,Guerrieri:2014nxa}). 

In this work we consider for the first time effects due to the spins of the heavy $\bar b$ quarks. In the first step of the Born-Oppenheimer approximation we compute the contribution of the light quarks $u d$ to the potential of the heavy antiquarks $\bar b \bar b$ via lattice QCD. As before \cite{Bicudo:2012qt,Bicudo:2015vta,Bicudo:2015kna} we use the static approximation for the $\bar b$ quarks, where their positions are frozen and their spin is irrelevant. Then, in the second step, we use a Hamiltonian for $\bar b$ quarks of finite mass with their spin interactions incorporated.

Notice the $\bar b$ spin effects are expected to be of the same order of magnitude as the estimated binding energy of the $u d \bar b \bar b$ tetraquark. For instance spin effects account for a mass difference $m_{B^\ast} - m_B \approx 46 \, \textrm{MeV}$, while the binding energy found in \cite{Bicudo:2015kna} is $E_B = -90^{+43}_{-36} \, \textrm{MeV}$. Moreover, both the kinetic term utilized in previous works, $p^2 / 2 \mu$ with $\mu = m_b/2$, and the spin-dependent part of the one-gluon exchange potential of a heavy and a light quark,
\begin{eqnarray}
\nonumber & & \hspace{-0.7cm} V_{j k}(\mathbf{r}_j,\mathbf{s}_j,\mathbf{r}_k,\mathbf{s}_k) \ \ = \ \ -\frac{C \alpha_s}{4} \bigg(\frac{1}{r} \\
\nonumber & & \hspace{0.675cm} - \frac{\pi}{2} \delta^3(\mathbf{r}) \left(\frac{1}{{m_j}^2} + \frac{1}{{m_k}^2} + \frac{16 \mathbf{s}_j \cdot \mathbf{s}_k}{3 m_j m_k}\right) + \ldots \bigg) , \\
\label{EQN650} & &
\end{eqnarray}
($j,k$ are the (anti)quark indices, $\mathbf{r}_j$, $\mathbf{s}_j$ and $m_j$ denote their positions, spins and masses, respectively; cf.\ \cite{DeRujula:1975qlm}) are of the same order in the $1/m_b$ expansion. Thus it is crucial to study the impact of the $\bar b$ spins on the predicted $u d \bar b \bar b$ tetraquark. 

At small $\bar b \bar b$ separations the heavy degrees of freedom are the two $\bar b$ quarks, while at large separations these are rather the two $B$ and/or $B^\ast$ mesons. Thus, in what concerns the kinetic energy, the reduced mass may either be $m_b / 2$ or $m_{B^{(\ast)}} / 2$. In Refs.\ \cite{Bicudo:2012qt,Bicudo:2015vta} we verified that both options result in very similar binding energies for the $u d \bar b \bar b$ tetraquark. Similar considerations apply to the heavy spin effects: at small separations we expect a hyperfine potential similar to (\ref{EQN650}), while at large separations heavy spin effects should be reflected by the mass difference $m_{B^*} - m_B$.

Hyperfine potentials for $u d \bar b \bar b$ systems from lattice QCD are not yet available, although heavy quark effective theory (HQET) could in principle be used to compute them (so far this has only been done for heavy quark-antiquark systems; cf.\ e.g.\ \cite{Koma:2006si,Koma:2006fw}). Thus we follow the strategy of including the heavy spin effects via the $m_{B^*}- m_B$ mass difference. We compute potentials of two heavy antiquarks $\bar b \bar b$ in the presence of two light $u d$ quarks for different light spin combinations using the static approximation for the $\bar b$ quarks \cite{Wagner:2010ad,Wagner:2011ev,Bicudo:2015kna}. Then we interpret them as potentials between appropriate linear combinations of pseudoscalar $B$ mesons and/or vector $B^*$ mesons as suggested in \cite{Scheunert:2015pqa}. These potentials, which correspond to $B B$,  $B B^*$ and $B^* B^*$ meson pairs, are finally used in a coupled system of non-relativistic Schrödinger equations for the relative coordinate of the two $\bar b$ quarks. This allows to investigate, how the binding energy of the $u d \bar b \bar b$ tetraquark is affected by the heavy spins. In particular we are able to check and will confirm, that the predicted $u d \bar b \bar b$ tetraquark with quantum numbers $I(J^P) = 0(1^+)$ still exists, when heavy spin effects are taken into account.

The paper is organized as follows. In section~\ref{SEC684} we discuss the Born-Oppenheimer approximation and include heavy spin effects by setting up a $16 \times 16$ coupled channel Schrödinger equation. In section~\ref{SEC379} we discuss symmetries and quantum numbers to split this equation into several independent problems, which are more easy to solve. In section~\ref{SEC322} we we discuss one of these problems in detail and solve it numerically, a $2 \times 2$ coupled channel Schr\"odinger equation corresponding to the $I(J^P) = 0(1^+)$ $u d \bar b \bar b$ tetraquark. Finally we conclude in section~\ref{SEC572}.


\section{\label{SEC684}Incorporating heavy $\bar{b} \bar{b}$ spin effects}


\subsection{\label{sec:relating operators}Interpreting lattice QCD $q q \bar{Q} \bar{Q}$ potentials in terms of $B$ and $B^\ast$ mesons}

In \cite{Bicudo:2015kna} we have computed potentials of two static antiquarks $\bar{Q}$ in the presence of two light quarks $q \in \{ u , d \}$ of physical mass using standard techniques from lattice QCD, which is the first step of the Born-Oppenheimer approximation. In particular we have used four-quark creation operators
\begin{eqnarray}
\nonumber & & \hspace{-0.7cm} \mathcal{O}_{L,S}(\vec{r}_1,\vec{r}_2) = (\mathcal{C} L)_{A B} (\mathcal{C} S)_{C D} \\
\label{EQN001} & & \hspace{0.675cm} \Big(\bar{Q}_C(\vec{r}_1) q_A^{(1)}(\vec{r}_1)\Big) \Big(\bar{Q}_D(\vec{r}_2) q_B^{(2)}(\vec{r}_2)\Big)
\end{eqnarray}
($\mathcal{C} = \gamma_0 \gamma_2$ is the charge conjugation matrix, $A,\ldots,D$ denote spin indices). The positions of the static quarks $\vec{r}_1$ and $\vec{r}_2$ are fixed, i.e.\ can be considered as quantum numbers. Moreover, the static quark spins do not appear in the Hamiltonian, i.e.\ the potentials do not depend on these static spins, and light and static spins are separately conserved. Thus, it is appropriate to couple the two light spins (via $L$) and the two static spins (via $S$). The corresponding potentials, which are independent of $S$, are denoted by $V_L(r)$, $r = |\vec{r}_1 - \vec{r}_2|$.

At large $r$ the considered $q q \bar{Q} \bar{Q}$ four-quark system will have the structure of two static-light mesons $\bar{Q} \Gamma q$ at separation $r$. Since a static quark has only two spin components, i.e.\ can also be denoted according to $\bar{Q} \rightarrow \bar{Q} (\mathds{1} + \gamma_0) / 2$, there are only 8 independent combinations of $\gamma$ matrices corresponding to the following quantum numbers:
\begin{itemize}
\item $\Gamma = (\mathds{1} + \gamma_0) \gamma_5$ \\
$\rightarrow$ $J^P = 0^-$ (the pseudoscalar $B$ meson),

\item $\Gamma = (\mathds{1} + \gamma_0) \gamma_j$ ($j = 1,2,3$) \\
$\rightarrow$ $J^P = 1^-$ (the vector $B^\ast$ meson),

\item $\Gamma = (\mathds{1} + \gamma_0) \mathds{1}$ \\
$\rightarrow$ $J^P = 0^+$ (the scalar $B_0^\ast$ meson),

\item $\Gamma = (\mathds{1} + \gamma_0) \gamma_j \gamma_5$ ($j = 1,2,3$) \\
$\rightarrow$ $J^P = 1^+$ (the pseudovector $B_1^\ast$ meson).
\end{itemize}
As already mentioned the static quark spins do not appear in the Hamiltonian and, hence, $B$ and $B^\ast$ mesons are degenerate as well as $B_0^\ast$ and $B_1^\ast$ mesons. For a comprehensive discussion of static-light mesons we refer to \cite{Jansen:2008si,Michael:2010aa}.

To understand the details of the meson-meson structure generated by the creation operators (\ref{EQN001}), one has to express them in terms of static-light bilinears $\bar{Q} \Gamma q$. We do this by using the Fierz identity,
\begin{eqnarray}
\nonumber & & \hspace{-0.7cm} \mathcal{O}_{L,S}(\vec{r}_1,\vec{r}_2) = \\
\label{eq:fierz identity} & & = \mathds{G}(S,L)_{a b} \Big(\bar{Q}(\vec{r}_1) \Gamma^{a} q^{(1)}(\vec{r}_1)\Big) \Big(\bar{Q}(\vec{r}_2) \Gamma^{b} q^{(2)}(\vec{r}_2)\Big)
\end{eqnarray}
with
\begin{eqnarray}
\mathds{G}(S,L)_{a b} = \frac{1}{16} \Tr \Big((\mathcal{C} S)^T \Gamma_a^T (\mathcal{C} L) \Gamma_b \Big) ,
\end{eqnarray}
where $\Gamma^a \in \{ (\mathds{1} + \gamma_0) \gamma_5 \, , \, (\mathds{1} + \gamma_0) \gamma_j \, , \, (\mathds{1} + \gamma_0) \mathds{1} \, , \, (\mathds{1} + \gamma_0) \gamma_j \gamma_5 \}$ (as discussed above) and $\Gamma_a$ denotes the inverse of $\Gamma^a$. From the right-hand side of (\ref{eq:fierz identity}) one can read off, which linear combinations of pairs of $B$, $B^\ast$, $B_0^\ast$ and $B_1^\ast$ mesons the creation operators $\mathcal{O}_{L,S}$ excite.

In this work we focus on combinations of pairs of $B$ and $B^\ast$ mesons (the two lightest bottom mesons), which are degenerate in the static limit and have similar mass in nature ($m_{B^\ast} - m_B \approx 45 \, \textrm{MeV}$). One can show that there are 16 posibilities of light and static spin couplings,
\begin{eqnarray}
L \, , \, S \in \{ (\mathds{1} + \gamma_0) \gamma_5 \, , \, (\mathds{1} + \gamma_0) \gamma_j \} ,
\end{eqnarray}
which generate exclusively such combinations, i.e.\ where $\mathds{G}(S,L)_{a b} = 0$, if either $\Gamma^a$ or $\Gamma^b$ in (\ref{eq:fierz identity}) is not element of $\{ (\mathds{1} + \gamma_0) \gamma_5 \, , \, (\mathds{1} + \gamma_0) \gamma_j \}$.

The corresponding $q q \bar{Q} \bar{Q}$ potentials, which depend only on $L$, but not on $S$, fall in two different classes,
\begin{itemize}
\item[(1)] $V_5(r) \equiv V_{(\mathds{1} + \gamma_0) \gamma_5}$,
\begin{itemize}
\item corresponding to $L = (\mathds{1} + \gamma_0) \gamma_5$,
\item attractive for isospin $I = 0$ \\ ($q q = (u d - d u) / \sqrt{2}$),
\item repulsive for isospin $I = 1$ \\ ($q q \in \{ uu \, , \, (u d + d u) / \sqrt{2} \, , \, dd \}$),
\end{itemize}

\item[(2)] $V_j(r) \equiv V_{(\mathds{1} + \gamma_0) \gamma_j}$,
\begin{itemize}
\item corresponding to $L = (\mathds{1} + \gamma_0) \gamma_j$,
\item repulsive for isospin $I = 0$ \\ ($q q = (u d - d u) / \sqrt{2}$),
\item attractive for isospin $I = 1$ \\ ($q q \in \{ uu \, , \, (u d + d u) / \sqrt{2} \, , \, dd \}$).
\end{itemize}
\end{itemize}

Note that neither for $V_5(r)$ (where $L = (\mathds{1} + \gamma_0) \gamma_5$) nor for $V_j(r)$ (where $L$ is an arbitrary linear combination of $(\mathds{1} + \gamma_0) \gamma_1 \, , \, (\mathds{1} + \gamma_0) \gamma_2 \, , \, (\mathds{1} + \gamma_0) \gamma_3$) it is possible to choose $S$ in such a way that exclusively a $B$ meson pair appears on the right hand side of Eq.\ (\ref{eq:fierz identity}) (i.e.\ $\Gamma^a = \Gamma^b = (\mathds{1} + \gamma_0) \gamma_5$). One always finds linear combinations of $B$ and $B^\ast$ mesons.

For example, when $L = S = (\mathds{1} + \gamma_0) \gamma_5$, the right-hand side of Eq.\ (\ref{eq:fierz identity}) is proportional to $B(\vec{r}_1) B(\vec{r}_2) + B_x^\ast(\vec{r}_1) B_x^\ast(\vec{r}_2) + B_y^\ast(\vec{r}_1) B_y^\ast(\vec{r}_2) + B_z^\ast(\vec{r}_1) B_z^\ast(\vec{r}_2)$ ($B \equiv \bar{Q} (\mathds{1} + \gamma_0) \gamma_5 q$ and $B_j^\ast \equiv \bar{Q} (\mathds{1} + \gamma_0) \gamma_j q$, i.e.\ $j = x,y,z$ denotes the spin orientation of $B^\ast$). Vice versa, a $B(\vec{r}_1) B(\vec{r}_2)$ pair does not have defined light quark spin and hence does not exclusively correspond to one of the two potentials $V_5(r)$ or $V_j(r)$, but to a mixture of both, where one is attractive and the other is repulsive.

Taking the mass difference and the mixing of $B$ and $B^\ast$ mesons into account (or the mixing of attractive and repulsive potentials, respectively), which has not been considered in our previous studies \cite{Bicudo:2012qt,Bicudo:2015vta,Bicudo:2015kna}, is the goal of this work.


\subsection{The coupled channel Schr\"odinger equation}

To determine, whether there are bound $u d \bar{b} \bar{b}$ states, we study a coupled channel Schr\"odinger equation for the two $\bar{b}$ quarks as the second step of the Born-Oppenheimer approximation,
\begin{eqnarray}
\label{eq:coupled SE} H \Psi(\vec{r}_1,\vec{r}_2) = E \Psi(\vec{r}_1,\vec{r}_2) .
\end{eqnarray}
The Hamiltonian $H$ acts on a 16-component wave function $\Psi$. The 16 components of $\Psi$ correspond to the 16 possibilities to combine $(B(\vec{r}_1) , B_x^\ast(\vec{r}_1) , B_y^\ast(\vec{r}_1) , B_z^\ast(\vec{r}_1))$ and $(B(\vec{r}_2) , B_x^\ast(\vec{r}_2) , B_y^\ast(\vec{r}_2) , B_z^\ast(\vec{r}_2))$, i.e.\ the first component corresponds to $B(\vec{r}_1) B(\vec{r}_2)$, the second to $B(\vec{r}_1) B_x^\ast(\vec{r}_2)$, the third to $B(\vec{r}_1) B_y^\ast(\vec{r}_2)$, etc. The Hamiltonian can be split in a free and an interacting part according to $H = H_0 + H_{\textrm{int}}$.

The free part of the Hamiltonian $H_0$ contains the kinetic energy of the $\bar{b}$ quarks and the masses of the $B$ and the $B^\ast$ mesons,
\begin{eqnarray}
H_0 =
M \otimes \mathds{1}_{4 \times 4} + \mathds{1}_{4 \times 4} \otimes M
+ \frac{\vec{p}_1^2 + \vec{p}_2^2}{2 m_b} \mathds{1}_{16 \times 16}
\end{eqnarray}
with
\begin{eqnarray}
M = \diag\Big(m_B \, , \, m_{B^\ast} \, , \, m_{B^\ast} \, , \, m_{B^\ast}\Big)
\end{eqnarray}
($m_b = 4977$ from the quark model \cite{PhysRevD.32.189} and $m_B = 5280 \, \textrm{MeV}$, $m_{B^\ast} = 5325 \, \textrm{MeV}$ from the PDG \cite{Agashe:2014kda}). It is illustrative to consider for the moment $H_{\textrm{int}} = 0$, i.e.\ the trivial case, where interactions between the $\bar{b}$ quarks are absent. Clearly, the system of 16 equations (\ref{eq:coupled SE}) decouples into 16 independent equations, the first for $B B$, the second for $B B_x^\ast$, the third for $B B_y^\ast$, etc. It is straightforward to determine the lowest energy eigenvalues of these 16 equations, which are $m_B + m_B$ ($1 \times$), $m_B + m_{B^\ast}$ ($6 \times$) and $m_{B^\ast} + m_{B^\ast}$ ($9 \times$), i.e.\ they correspond to the sum of the two corresponding non-interacting mesons.

The interacting part of the Hamiltonian $H_{\textrm{int}}$ contains the $\bar{Q} \bar{Q} q q$ potentials $V_5(r)$ and $V_j(r)$ computed with lattice QCD and discussed in the previous subsection. These potentials are spherically symmetric and can be parameterized by
\begin{eqnarray}
\label{eq:fit function}	V_X(r) = -\frac{\alpha_X}{r} \exp\bigg(-\bigg(\frac{r}{d_X}\bigg)^2\bigg) ,
\end{eqnarray}
where $\alpha_X$ and $d_X$ ($X = 5, j$) are determined by fitting (\ref{eq:fit function}) to our lattice QCD results from \cite{Bicudo:2015kna}. At small $\bar{b} \bar{b}$ separations the potentials are dominated by 1-gluon exchange and, hence, are proportional to $1/r$, while at large $\bar{b} \bar{b}$ separations there is exponential screening, which corresponds to the formation of an essentially non-interacting $B^{(\ast)}B^{(\ast)}$ meson pair (for details cf.\ \cite{Bicudo:2015vta,Bicudo:2015kna}). $H_{\textrm{int}}$ is given by
\begin{eqnarray}
H_{\textrm{int}} = T^{-1} V(r) T ,
\end{eqnarray}
where
\begin{eqnarray}
\nonumber  & & \hspace{-0.7cm} V(r) = \diag\Big(\underbrace{V_5(r),\ldots,V_5(r)}_{4 \times} , \underbrace{V_j(r),\ldots,V_j(r)}_{12 \times}\Big) . \\
 & &
\end{eqnarray}
$T$ is the transformation between the 16 components of the Schr\"odinger equation i.e.\ the 16 possible meson pairs $B B$, $B B_x^\ast$, $B B_y^\ast$, etc.\ and the 16 static-static-light-light channels defined by the static and the light spin couplings $S$ and $L$ (cf.\ Eq.\ (\ref{EQN001})), for which the $q q \bar{Q} \bar{Q}$ potentials have been computed. The entries of $T$ are the coefficients $\mathds{G}(S,L)_{a b}$ appearing in the Fierz identity (\ref{eq:fierz identity}), where $S,L$ label the rows and $a b$ label the columns of $T$. $T$ is not diagonal and, hence, couples the 16 equations (\ref{eq:coupled SE}). The corresponding physics is the interplay between different meson masses $m_B$ and $m_{B^\ast}$ on the one hand and attractive and repulsive potentials $V_5(r)$ and $V_j(r)$ on the other hand.


\section{\label{SEC379}Symmetries and Quantum numbers}


\subsection{Decoupling the Schr\"odinger equation according to total angular momentum}

The $16 \times 16$ coupled channel Schr\"odinger equation (\ref{eq:coupled SE}) can be decoupled into simpler $1 \times 1$ or $2 \times 2$ equations, which correspond to total angular momentum $J = 0,1,2$, $J_z = -J,\ldots,+J$ and, in case of $J = 1$, to symmetry/antisymmetry with respect to meson exchange.


\subsubsection{Total angular momentum $J = 0$}

For $J=0$ there is a single $2 \times 2$ coupled channel Schr\"odinger equation with Hamiltonian, 
\begin{subequations}
\label{eq:HJ0}
\begin{align}	
& \tilde{H}_{0, J=0} =
\begin{pmatrix}
2 m_B & 0 \\
0 & 2 m_{B^{\ast}}
\end{pmatrix}
+ \frac{\vec{p}_1^2 + \vec{p}_2^2}{2 m_b} \mathds{1}_{2 \times 2} \\
& \tilde{H}_{\textrm{int}, J=0} = \frac{1}{4}
\begin{pmatrix}
V_5(r) + 3 V_j(r) & \sqrt{3} (V_5(r) - V_j(r)) \\
\sqrt{3} (V_5(r) - V_j(r)) & 3 V_5(r) + V_j(r) 
\end{pmatrix} .
\end{align}	
\end{subequations}
The corresponding 2-component wave function is related to the components of the 16-component wave function from (\ref{eq:coupled SE}) via,
\begin{equation}
\tilde{\Psi}_{J=0} =
\begin{pmatrix}
B B \\
(1 / \sqrt{3}) (\vec{B}^{\ast})^2
\end{pmatrix}
\end{equation}
with $(\vec{B}^{\ast})^2 = B^{\ast}_x B^{\ast}_x + B^{\ast}_y B^{\ast}_y + B^{\ast}_z B^{\ast}_z$.


\subsubsection{Total angular momentum $J = 1$}

For $J = 1$ there is a threefold degeneracy (due to $J_z = -1,0,+1$) both for a $1 \times 1$ Schr\"odinger equation and a $2 \times 2$ Schr\"odinger equation.
\begin{itemize}
\item The Hamiltonian of each of the three $1 \times 1$ equations is
\begin{subequations}
\label{eq:HJ11x1}		
\begin{align}
& \tilde{H}_{0, J=1, 1 \times 1} = m_B + m_{B^\ast} + \frac{\vec{p}_1^2 + \vec{p}_2^2}{2 m_b} \\
& \tilde{H}_{\textrm{int}, J=1, 1 \times 1} = V_j(r)
\end{align}
\end{subequations}
and the corresponding wave functions are symmetric under meson exchange,
\begin{eqnarray}
\tilde{\Psi}_{J=1, j, 1 \times 1} = \frac{1}{\sqrt{2}} \Big(B_j^\ast B + B B_j^\ast\Big) .
\end{eqnarray}

\item The Hamiltonian of each of the three $2 \times 2$ equations is
\begin{subequations}
\label{eq:HJ12x2}
\begin{align}
& \tilde{H}_{0, J=1, 2 \times 2} =
\begin{pmatrix}
m_{B^\ast} + m_B & 0 \\
0 & 2 m_{B^\ast}
\end{pmatrix}
+ \frac{\vec{p}_1^2 + \vec{p}_2^2}{2 m_b} \mathds{1}_{2 \times 2} \\
& \tilde{H}_{\textrm{int}, J=1, 2 \times 2} = \frac{1}{2}
\begin{pmatrix}
V_5(r) + V_j(r) & V_j(r) - V_5(r) \\
V_j(r) - V_5(r) & V_5(r) + V_j(r)
\end{pmatrix}
\end{align}
\end{subequations}
and the corresponding 2-component wave functions are antisymmetric under meson exchange,
\begin{equation}
\tilde{\Psi}_{J=1, j, 2 \times 2} = \frac{1}{\sqrt{2}}
\begin{pmatrix}
B_j^\ast B - B B_j^\ast \\
\epsilon_{j k l} B_k^\ast B_l^\ast
\end{pmatrix} .
\end{equation}			
\end{itemize}


\subsubsection{Total angular momentum $J = 2$}

For $J = 2$ there is a fivefold degeneracy (due to $J_z = -2,-1,0,+1,+2$) for a $1 \times 1$ Schr\"odinger equation. The Hamiltonian of each of the five $1 \times 1$ equations is
\begin{subequations}
\label{eq:HJ2}
\begin{align}
& \tilde{H}_{0, J=2, 1 \times 1} = 2 m_{B^\ast} + \frac{\vec{p}_1^2 + \vec{p}_2^2}{2 m_b} \\
& \tilde{H}_{\textrm{int}, J=2, 1 \times 1} = V_j(r)
\end{align}
\end{subequations}
with corresponding wave functions
\begin{eqnarray}
\tilde{\Psi}_{J=2, J_z} = T_{2, J_z}(B_x^\ast , B_y^\ast , B_z^\ast) ,
\end{eqnarray}
where $T_{2, J_z}$ are the components of a spherical tensor of rank 2, which is quadratic in $B_x^\ast$, $B_y^\ast$ and $B_z^\ast$.


\subsection{\label{SEC548}Antisymmetry of the wave function and isospin}

Each of the four Hamiltonians (\ref{eq:HJ0}), (\ref{eq:HJ11x1}), (\ref{eq:HJ12x2}) and (\ref{eq:HJ2}) is valid only for either isospin $I = 0$ (where $V_5$ attractive and $V_j$ is repulsive) or isospin $I = 1$ (where $V_5$ is repulsive and $V_j$ is attractive). The reason is that both the heavy antiquarks $\bar{b} \bar{b}$ as well as the light quarks $q q$ are fermions and, therefore, their wave function must be antisymmetric under exchange according to the Pauli exclusion principle. For the $\bar{b}$ quarks this is neglected in our lattice QCD computations of the potentials, since we have used static quarks.

In \cite{Bicudo:2015vta} the quantum numbers of the heavy antiquarks $\bar{b} \bar{b}$ and the light quarks $q q$ have been discussed in detail to explain, why certain potentials are attractive, while others are repulsive. Here we summarize these arguments again and relate isospin $I$, light spin $j$ and heavy spin $j_b$ to the Hamiltonians (\ref{eq:HJ0}), (\ref{eq:HJ2}), (\ref{eq:HJ11x1}) and (\ref{eq:HJ12x2}) characterized by total angular momentum $J$ (cf.\ Table~\ref{tab:symmetries} for a summary).

\begin{table*}[htb]
	\centering
	\begin{tabular}{|c||c|c|c||c|c||c|}
		\hline
                & \multicolumn{3}{c||}{} & \multicolumn{2}{c||}{} & \vspace{-0.3cm} \\
		& \multicolumn{3}{c||}{light quarks $q q$} & \multicolumn{2}{c||}{heavy antiquarks $\bar{b}\bar{b}$} & $q q \bar{b} \bar{b}$
		\\
                & \multicolumn{3}{c||}{} & \multicolumn{2}{c||}{} & \vspace{-0.3cm} \\
		\hline
                & & & & & & \vspace{-0.3cm} \\
		combination & isospin $I$ & spin $j$ & color & color & spin $j_b$ & spin, parity $J^P$ \\  
                & $\quad \quad \quad \quad$ & $\quad \quad \quad \quad$ & $\quad \quad \quad \quad$ & $\quad \quad \quad \quad$ & $\quad \quad \quad \quad$ & \vspace{-0.3cm} \\
		\hline
                & & & & & & \vspace{-0.3cm} \\
		1 & $0$ (A) & $0$ (A) & $\bar{3}$ (A) & $3$ (A) & $1$ (S) & $1^+$ \\  
		2 & $0$ (A) & $1$ (S) & $6$ (S) & $\bar{6}$ (S) & $0$ (A) & $1^+$ \\  
                & & & & & & \vspace{-0.3cm} \\
		\hline
                & & & & & & \vspace{-0.3cm} \\
		3 & $1$ (S) & $0$ (A) & $6$ (S) & $\bar{6}$ (S) & $0$ (A) & $0^+$ \\  
		4 & $1$ (S) & $1$ (S) & $\bar{3}$ (A) & $3$ (A) & $1$ (S) & $0^+$, $1^+$, $2^+$ \vspace{-0.3cm} \\  
                & & & & & & \\
		\hline
	\end{tabular}
	\caption{\label{tab:symmetries}Possible combinations of quantum numbers/color representations and corresponding symmetric (S) or antisymmetric (A) behavior of the wave functions.}
\end{table*}
\renewcommand{\arraystretch}{1}

The computed potentials of two static antiquarks $V_L$ appearing in (\ref{eq:HJ0}), (\ref{eq:HJ11x1}), (\ref{eq:HJ12x2}) and (\ref{eq:HJ2}) are different for different isospin $I = 0,1$ and different light spin $j = 0,1$ ($L = (\mathds{1} + \gamma_0) \gamma_5$ corresponds to $j = 0$, $L = (\mathds{1} + \gamma_0) \gamma_j$ corresponds to $j = 1$). Therefore, we will discuss four possibilities (the four lines of Table~\ref{tab:symmetries}). 

If isospin and light spin are identical, i.e.\ either $I = j = 0$ (combination~1) or $I = j = 1$ (combination~4), the light quarks must be in an antisymmetric color triplet $\bar{3}$. For a gauge invariant four-quark system this implies also an antisymmetric color triplet $3$ for the $\bar{b}$ quarks, i.e.\ an attractive potential. Similarly, if isospin and light spin are not identical, i.e.\ either $I = 0 \neq j = 1$ (combination~2) or $I = 1 \neq j = 0$ (combination~3), both the light and the heavy quarks must be in a repulsive color sextet $6$ and $\bar{6}$, respectively, i.e.\ the potential is repulsive.

We expect that for a possibly existing bound state the $\bar{b}$ quarks form a spatially symmetric s-wave (in section~\ref{SEC322} we solve the Schr\"odinger equation for such an s-wave). Consequently, the heavy spin $j_b$ must be symmetric for color triplets, i.e.\ $j_b = 1$, and antisymmetric for color sextets, i.e.\ $j_b = 0$.

The $I = 0$ combinations 1 and 2 both have total angular momentum $J = 1$ (either $j = 0$, $j_b = 1$ or $j = 1$, $j_b = 0$). They correspond to the $2 \times 2$ problem (\ref{eq:HJ12x2}) with an attractive $V_5$ and a repulsive $V_j$ potential. In our previous papers \cite{Bicudo:2012qt,Bicudo:2015vta,Bicudo:2015kna}, where we did not take heavy spin effects into account, we found that this attactive $V_5$ potential is sufficiently strong to host a bound state, which was interpreted as an $I(J^P) = 0(1^+)$ tetraquark. The main question of this paper, which we will investigate in the next section, is, whether this bound state will persist, when we consider the $2 \times 2$ problem (\ref{eq:HJ12x2}), i.e.\ when including heavy spin effects.

The $I = 1$ combinations 3 and 4 have $J = 0$ and $J = 0, 1, 2$, respectively. They correspond to the $2 \times 2$ problem (\ref{eq:HJ0}) and the $1 \times 1$ problems (\ref{eq:HJ11x1}) and (\ref{eq:HJ2}) with an attractive $V_j$ and a repulsive $V_5$ potential. Since the $I = 1$ $V_j$ potential has been found to be not sufficiently attractive to host a bound state, even with heavy spin effects neglected, we do not study these combinations any further.

Since both $B$ and $B^{\star}$ mesons have negative parity, combinations 1 to 4 all have positive parity $P = +$.


\section{\label{SEC322}Solving the coupled Schrödinger equation}


\subsection{Analytical simplifications and boundary conditions}

The $2 \times 2$ coupled channel Schrödinger equation (\ref{eq:HJ12x2}) is a partial differential equation in six variables, the positions of the $\bar{b}$ quarks $\vec{r}_1$ and $\vec{r}_2$. It can be split in two independent equations in three variables by transforming to the center of mass coordinate (the corresponding equation is trivial to solve) and the relative coordinate $\vec{r} = \vec{r}_2 - \vec{r}_1$.

Since the potentials $V_5$ and $V_j$ are spherically symmetric, the Schrödinger equation for the relative coordinate reduces to an ordinary differential equation in the variable $r = |\vec{r}|$ and its solutions have defined orbital angular momentum. As discussed in section~\ref{SEC548} we study $\bar{b} \bar{b}$ in an s-wave,
\begin{eqnarray}
\nonumber & & \hspace{-0.7cm}
\bigg(
\begin{pmatrix}
m_{B^\ast} + m_B & 0 \\
0 & 2 m_{B^\ast}
\end{pmatrix}
- \frac{\hbar}{2 \mu} \frac{\mathrm{d}^2}{\mathrm{d}r^2} \mathds{1}_{2 \times 2} \\
\nonumber & & \hspace{0.675cm} +
\begin{pmatrix}
V_j(r) + V_5(r) & V_j(r) - V_5(r) \\
V_j(r) - V_5(r) & V_j(r) + V_5(r)
\end{pmatrix}
\bigg) \chi(r) = \\
\label{eq:radSG} & & = E \chi(r) ,
\end{eqnarray} 
where $\mu = m_b / 2$ and
\begin{eqnarray}
\label{EQN764} \chi(r) =
\begin{pmatrix}
\chi_1 (r) \\
\chi_2 (r)
\end{pmatrix}
= \psi(r) r
\end{eqnarray}
with $\psi$ denoting the wave function of the relative coordinate $\vec{r}$.

Generalizing well-known results from quantum mechanics to our $2 \times 2$ coupled channel Schrödinger equation we find that $\chi(r)$ has to vanish linearly for small $r$, i.e.\
\begin{eqnarray}
\label{eq:asymptoticBehavior} \chi(r) \sim \begin{pmatrix} A r \\ B r \end{pmatrix} \quad \textrm{for} \quad r \rightarrow 0 .
\end{eqnarray}
Moreover, if $\chi(r)$ describes a bound state, it has to vanish exponentially for large $r$, i.e.\
\begin{eqnarray}
\label{eq:infiniteBehavior} \chi(r) \sim
\begin{pmatrix} C e^{-\Delta E r} \\ D e^{-\Delta E r} \end{pmatrix}
\quad \textrm{for} \quad r \rightarrow \infty
\end{eqnarray}
with $\Delta E = m_B + m_{B^\ast} - E$.


\subsection{\label{sec:numercial solution}Numerical solution}

We solve Eq.\ (\ref{eq:radSG}) for boundary conditions (\ref{eq:asymptoticBehavior}) and (\ref{eq:infiniteBehavior}) numerically by employing the shooting method. The shooting method is an iterative root finding procedure, where in each step the ordinary differential equation (\ref{eq:radSG}) has to be solved, which we do by using the Runge-Kutta-Fehlberg method.

One possibility is to start the Runge-Kutta-Fehlberg computation of $\chi(r)$ at very small, but non-vanishing $r > 0$ ($r = 0$ would cause numerical problems due to the singularities of $V_5(r)$ and $V_j(r)$ at $r = 0$) with boundary conditions (\ref{eq:asymptoticBehavior}). The resulting $\chi(r)$ depends on two parameters, $E$ and $A/B$ (the absolute values of $A$ and $B$ only affect the normalization of the wave function $\chi(r)$ and, hence, are irrelevant). The values of the components of $\chi(r)$ at $r = r_{\max} \gtrsim 20 \, \textrm{fm}$, $\chi_1 (r_{\max})$ and $\chi_2 (r_{\max})$, are then used as input for a two-dimensional root solver, to determine the parameters $E$ and $A / B$ such that also the boundary conditions $\chi_1 (r_{\max}) = \chi_2 (r_{\max}) = 0$ (i.e.\ Eq.\ (\ref{eq:infiniteBehavior})) are fulfilled. In case a zero is found, the resulting $E < m_B + m_{B^\ast}$ is the energy of a bound state, i.e.\ the mass of a $u d \bar{b} \bar{b}$ tetraquark. In practice, however, multi-dimensional root finding is a non-trivial numerical problem. For example, we observed many cases, where each of the multi-dimensional root finding algorithms implemented in the GNU Scientific Library \cite{Gough:2009} failed to converge, even when the initial shooting values of the parameters $E$ and $A/B$ were chosen close to a solution.

Therefore, we resort to a more powerful and efficient variant of the shooting method (cf.\ e.g.\ \cite{Amann,Bicudo:1989si,Berwein:2015vca}), where root finding is reduced to only one dimension, the energy eigenvalue $E$. Let $\chi^{(A)}(r)$ and $\chi^{(B)}(r)$ denote solutions of (\ref{eq:radSG}) with asymptotic behavior
\begin{eqnarray}
\label{eq:specAsymBehavior} & & \hspace{-0.7cm} \chi^{(A)}(r) \sim
\begin{pmatrix} r \\ 0 \end{pmatrix}
\quad \textrm{for} \quad r \rightarrow 0 \\
 & & \hspace{-0.7cm} \chi^{(B)}(r) \sim
\begin{pmatrix} 0 \\ r \end{pmatrix}
\quad \textrm{for} \quad r \rightarrow 0 .
\end{eqnarray}
Note that both solutions are consistent with the boundary conditions (\ref{eq:asymptoticBehavior}) at $r \rightarrow 0$, but neither of them fulfills the boundary conditions (\ref{eq:infiniteBehavior}) at $r = r_{\max}$. Since the differential equation (\ref{eq:radSG}) is linear, one can combine its solutions $\chi^{(A)}(r)$ and $\chi^{(B)}(r)$ to a more general solution $\chi(r) = A \chi^{(A)}(r) + B \chi^{(B)}(r)$. This solution is still consistent with the boundary conditions (\ref{eq:asymptoticBehavior}) and has also the potential to fulfill the boundary conditions (\ref{eq:infiniteBehavior}), 
\begin{eqnarray}
\label{eq:linearSystemAB} \chi(r_{\max}) = A \chi^{(A)}(r_{\max}) + B \chi^{(B)}(r_{\max}) =
\begin{pmatrix} 0 \\ 0 \end{pmatrix} ,
\end{eqnarray}
for appropriately chosen $E$, $A$ and $B$. We exclude the trivial and physically not interestimg solution $A = B = 0$. There are additional non-trivial solution, if $\chi^{(A)}(r_{\max})$ and $\chi^{(B)}(r_{\max})$ are linerly dependent, i.e.\ if
\begin{equation}
\label{eq:detZero} \det\left(
\begin{array}{cc}
\chi_1^{(A)}(r_{\max}) & \chi_1^{(B)}(r_{\max}) \\ 
\chi_2^{(A)}(r_{\max}) & \chi_2^{(B)}(r_{\max})
\end{array}
\right) = 0 .
\end{equation}
Since the left hand side of (\ref{eq:detZero}) depends on $E$, but neither on $A$ nor on $B$, a simple one-dimensional root finding algorithm is sufficient. As soon as a solution $E$ is found, one can obtain $A/B$ via $A/B = -\chi_1^{(B)}(r_{\max}) / \chi_1^{(A)}(r_{\max}) = -\chi_2^{(B)}(r_{\max}) / \chi_2^{(A)}(r_{\max})$.


\subsection{Results}

The lattice QCD computation of the potentials $V_5(r)$ and $V_j(r)$ is explained in detail in \cite{Bicudo:2015kna}. We have used 2-flavor gauge link configurations generated by the European Twisted Mass Collabortaion (ETMC) \cite{Boucaud:2007uk,Boucaud:2008xu}.

For the attractive $I = 0$ potential $V_5(r)$ we have performed identical computations on several different ensembles to extrapolate the potential to physically light $u/d$ quark masses. Moreover, an evolved procedure to compute statistical errors and to estimate systematic errors has been applied. The lattice QCD results are consistently described by (\ref{eq:fit function}) with parameters $\alpha_5 = 0.34^{+0.03}_{-0.03}$, $d_5 = 0.45^{+0.12}_{-0.10} \, \textrm{fm}$ (for details cf.\ \cite{Bicudo:2015kna}).

The repulsive $I = 0$ potential $V_j(r)$ has been computed in the same way. Statistical errors are, however, much larger, such that a precise and stable quark mass extrapolation is not possible. A consistent parametrization of the lattice QCD results is again given by (\ref{eq:fit function}) with parameters $\alpha_j = 0.10 \pm 0.07$, $d_5 = (0.28 \pm 0.17) \, \textrm{fm}$, where the errors have been estimated in a rather crude but conservative way.

\begin{figure}[t]
\centerline{\includegraphics[width=0.95\columnwidth]{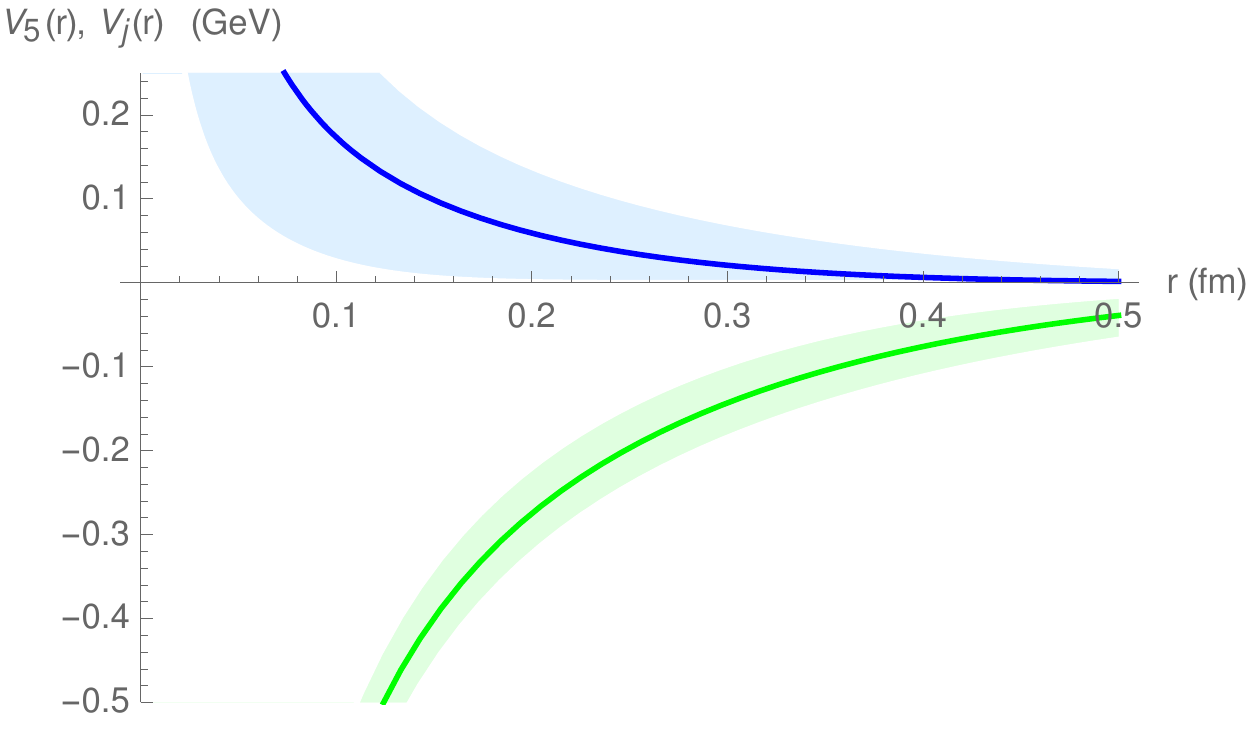}}
\caption{\label{FIG001}The attractive $I = 0$ potential $V_5(r)$ (green) and the repulsive $I = 0$ potential $V_j(r)$ (blue); the error bands reflect the uncertainties of the parameters $\alpha_5$, $d_5$, $\alpha_j$ and $d_j$ provided in the text.}
\end{figure}

In Figure~\ref{FIG001} both potentials are shown with corresponding error bands. These potentials are then used in the $2 \times 2$ coupled channel Schr\"odinger equation (\ref{eq:radSG}). We observe that our results concerning the mass $E$ and the binding energy $m_B + m_{B^\ast} - E$ of the $u d \bar b \bar b$ tetraquark strongly depend on $V_5(r)$ (which is available rather precisely), but are essentially independent of $V_j(r)$ (e.g.\ varying $\alpha_j$ and $d_j$ by $\pm 50\%$ does not change the results). In particular we find that the bound state in the $I(J^P) = 0(1^+)$ channel persists. The binding energy is
\begin{eqnarray}
\Delta E = m_B + m_{B^\ast} - E = 59_{-38}^{+30} \, \textrm{MeV} ,
\end{eqnarray}
i.e.\ we confirm the existence of the $u d \bar b \bar b$ tetraquark with a confidence level of nearly $2 \, \sigma$. Its mass is
%
%
\begin{eqnarray}
E = 10545_{-30}^{+38} \, \textrm{MeV}
\end{eqnarray}
(for comparison, in previous work, where we did not consider the heavy $\bar b$ spins, we found $\Delta E = 90_{-36}^{+43} \, \textrm{MeV}$ \cite{Bicudo:2015kna}, i.e., as expected and discussed in section~\ref{SEC684}, heavy spin effects reduce the binding energy.). The errors for both the binding energy $\Delta E$ and the mass $E$, are based on many jackknife samples already generated during the lattice QCD computation of the potential $V_5(r)$, e.g.\ correlations between $\alpha_5$ and $d_5$ are taken into account. Moreover, systematic errors are also included, e.g.\ by varying the temporal fitting range to extract the potentials $V_5(r)$ and $V_j(r)$. For details regarding the computation of these errors we refer to \cite{Bicudo:2015kna}.

\begin{figure}[b]
\centerline{\includegraphics[width=0.95\columnwidth]{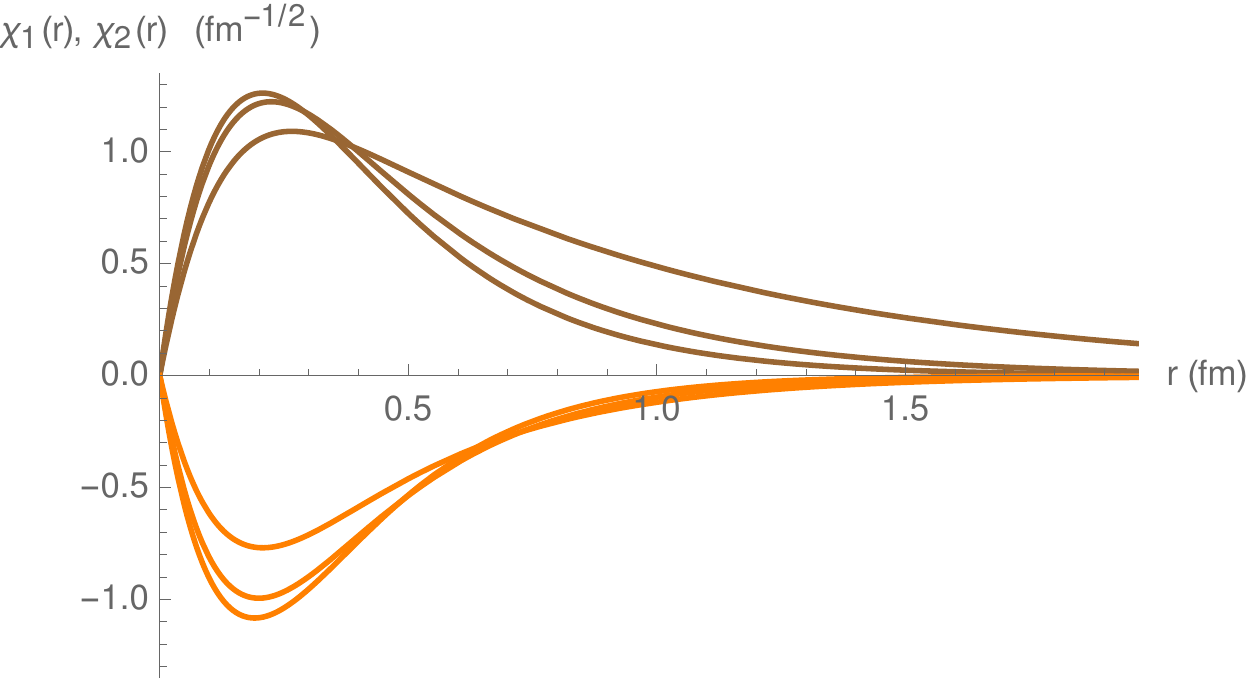}}
\caption{\label{figt:two_wfs}The two components of the wave function (\ref{EQN764}), $\chi_1(r)$ (the $B B^\ast$ component, brown) and $\chi_2(r)$ (the $B^\ast B^\ast$ component, orange); the three curves reflect the uncertainties of the parameters $\alpha_5$ and $d_5$ provided in the text.}
\end{figure}

In Figure~\ref{figt:two_wfs} we show the two components of the wave function (\ref{EQN764}), $\chi_1(r)$ (the $B B^\ast$ component, brown curves) and $\chi_2(r)$ (the $B^\ast B^\ast$ component, orange curves). For small $r \ltapprox 0.3 \, \textrm{fm}$ both components are of similar magnitude. In other words, a roughly even mixture of $B B^\ast$ and the heavier $B^\ast B^\ast$ is energetically preferred, since it corresponds to a good approximation to the potential $V_5(r)$, which is strongly attractive for small $r$. On the other hand, for large $r \gtapprox 0.6 \, \textrm{fm}$ the behavior is different, $|\chi_1(r)| \gg |\chi_2(r)|$, i.e.\ when the potentials $V_5(r)$ and $V_j(r)$ become weaker, the lighter $B B^\ast$ structure is favored.

\begin{figure}[htb]
\centerline{\includegraphics[width=0.95\columnwidth]{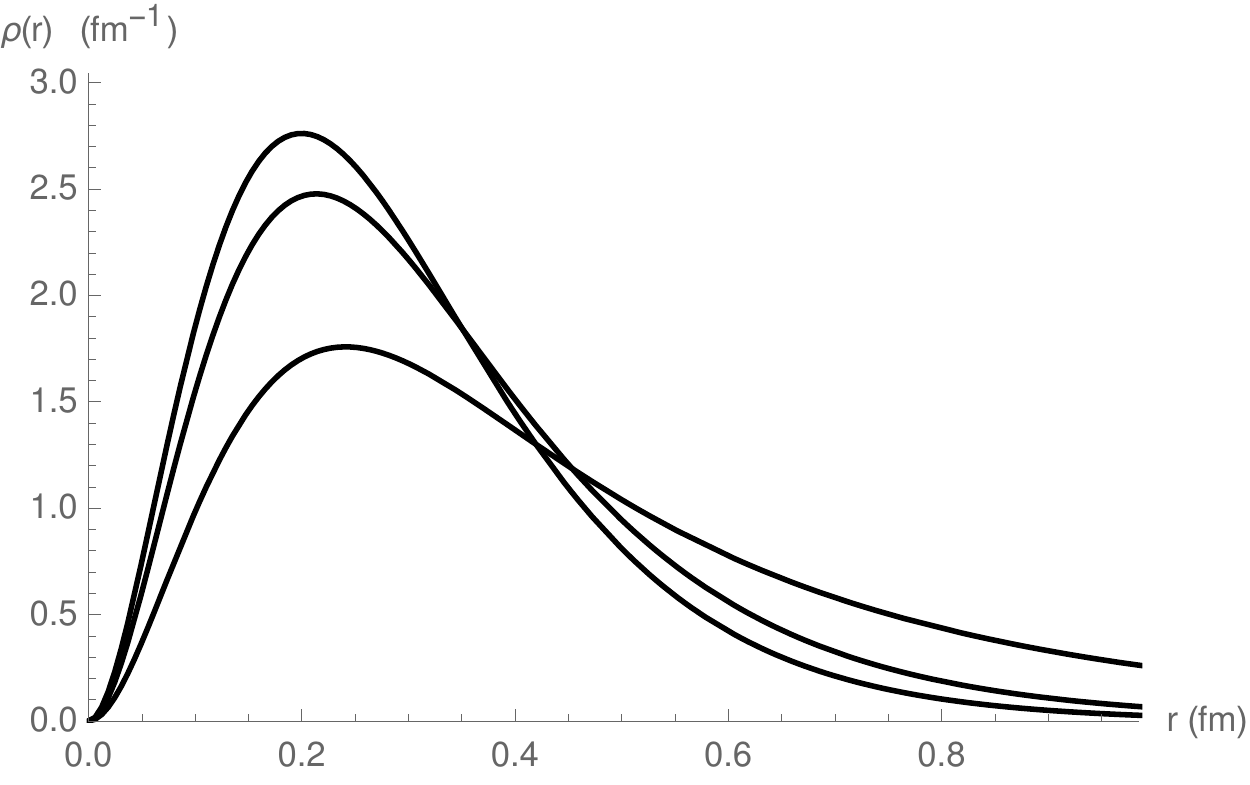}}
\caption{\label{fig:probability_density}The radial probability density $\rho(r) = |\chi_1(r)|^2 + |\chi_2(r)|^2$ for the separation of the heavy $\bar b$ quarks; the three curves reflect the uncertainties of the parameters $\alpha_5$ and $d_5$ provided in the text.}
\end{figure}

In Figure~\ref{fig:probability_density} we show the radial probability density for the separation of the heavy $\bar b$ quarks. One can see that a measurement of the $\bar b \bar b$ separation will typically result in a value $0.1 \, \textrm{fm} \ldots 0.5 \, \textrm{fm}$. In this region both components of the wave function $\chi(r)$ are sizable, as discussed in the previous paragraph (cf.\ also Figure~\ref{figt:two_wfs}). The conclusion is that the predicted $u d \bar b \bar b$ tetraquark is not just a combination of the two lightest mesons, i.e.\ $B$ and $B^\ast$, as one might naively expect. It is rather a linear superposition of a $B B^\ast$ and a $B^\ast B^\ast$ structure, where the latter is quite significant. This should be of particular interest e.g.\ for lattice computations of $u d \bar b \bar b$ systems using four quarks of finite mass, where suitable creation operators need to be chosen (cf.\ e.g.\ \cite{Francis:2016hui,Peters:2016isf}), or for corresponding Dyson-Schwinger Bethe-Salpeter studies (cf.\ e.g.\ \cite{Heupel:2012ua,Eichmann:2015cra}, where, however, flavor combinations different from $u d \bar b \bar b$ have been considered).

\section{\label{SEC572}Conclusions}

We have studied the effects of the heavy $\bar b$ quark spins on $u d \bar b \bar b$ tetraquark binding, using static-static-light-light potentials computed with lattice QCD and the Born-Oppenheimer approximation. We also utilize as input the masses of the pseudoscalar $B$ meson and vector $B^*$ meson. 

We simplify and block diagonalize the resulting large system of 16 coupled Schrödinger equations with the help of the Fierz identity and according to irreducible total angular momentum representations. We find that only one of the resulting blocks is a candidate for a bound four-quark state, which corresponds to our previously predicted $u d \bar b \bar b$ tetraquark with quantum numbers $I(J^P) = 0(1^+)$ \cite{Bicudo:2012qt,Bicudo:2015vta,Bicudo:2015kna}. In this block two channels are coupled, a $B B^*$ pair and a $B^* B^*$ pair.

Solving the corresponding coupled channel Schrödinger equation numerically we find that the spin of the heavy $\bar b$ quarks decreases the binding energy. Nevertheless, the attraction is sufficiently strong such that the bound four-quark state persists. Thus, we confirm our previous results and present even stronger evidence for an exotic tetraquark with flavour $u d \bar b \bar b$, isospin $I = 0$, total angular momentum $J = 1$ and parity $P = +$. It is a boundstate with respect to QCD with a mass, which is $\Delta E = 59_{-38}^{+30} \, \textrm{MeV}$ below the $B B^*$ threshold, but may decay due to the weak interactions.


\acknowledgements

We thank A.~Peters for providing static-static-light-light potentials computed with lattice QCD. We acknowledge useful conversations with K.~Cichy, C.~Fischer and A.~Peters.

P.B. acknowledges the support of CFTP (grant FCT UID/FIS/00777/2013) and is thankful for hospitality at the Institute of Theoretical Physics of Johann Wolfgang Goethe-University Frankfurt am Main. M.W.\ acknowledges support by the Emmy Noether Programme of the DFG (German Research Foundation), grant WA 3000/1-1.

This work was supported in part by the Helmholtz International Center for FAIR within the framework of the LOEWE program launched by the State of Hesse.

Calculations on the LOEWE-CSC and on the on the FUCHS-CSC high-performance computer of the Frankfurt University were conducted for this research. We would like to thank HPC-Hessen, funded by the State Ministry of Higher Education, Research and the Arts, for programming advice.


\bibliographystyle{apsrev4-1}
\bibliography{literature.bib}


\end{document}